\documentstyle[prl,multicol,aps,epsf]{revtex}
\begin{document}
\title{Mesoscopic mechanism of adiabatic charge transport}
\author{F. Zhou$^{a}$, B. Spivak$^{b}$, B. Altshuler$^{a,c}$}
\address{$^{a}$Physics Department, Princeton University, Princeton, NJ 08544}
\address{$^{b}$Physics Department, University of Washington, Seattle, WA 98195}
\address{$^{c}$NEC Research Institute, 4 Independence Way, Princeton, NJ 08540}
\maketitle

\begin{abstract}
We consider adiabatic charge transport through mesoscopic metallic
samples caused by a periodically
changing external potential. 
We find that both the amplitude and the sign of the charge transferred
through a sample per period are random sample specific quantities.
The characteristic
magnitude of the charge is determined by the quantum
interference.  
\end{abstract}
\begin{multicols}{2}
Let us apply an external potential
$\phi({\bf r},t)$, which is changing slowly and {\em periodically} in time 
to a metallic sample. 
This potential causes finite
net charge $Q$ transported across the sample per period.
This phenomenon known as adiabatic charge transport[1], 
has been investigated in 
several papers[1-4] 
for closed systems at zero temperature.  In
the adiabatic approximation, quantum state of such a system 
is characterized by its 
ground state wave function corresponding to the instantaneous value of  
the external potential $\phi({\bf r}, t)$. 
However, in real
experimental situations exact eigenfunctions of electrons are ill
defined: the electron energy levels are broadened 
due to inelastic processes at $T \neq 0$, and, in the case of
an open system, are further broadened due to finite dwell time.

In this article we present a theory of adiabatic charge transport in
 mesoscopic systems with "open geometries". We demonstrate that at
low $T$  both the magnitude and the sign of 
$Q$ are {\em sample specific} quantities. 
The typical value of
$Q$ in disordered (chaotic) systems turns out to be determined by quantum
interference effects. We evaluate this value and find that it is
much larger than the one in ballistic systems.
This enhancement manifests of the well known fact,
that at low temperatures all electronic characteristics of
mesoscopic samples are extremely sensitive to changes in the scattering
potential[5-8].

Let us start with a qualitative picture of the 
mesoscopic adiabatic charge transport.
The wave functions of electrons in mesoscopic disordered systems 
are determined by the particular realization of 
an impurity potential
and exhibit sample specific spatial fluctuations. 
Therefore, the spatial electron density profile
is changing slowly in time, 
together with the external potential $\phi({\bf r}, t)$.
According to the continuity relation, variation of the charge
density in time requires currents in the system.
The question arises:
What is the condition for a total charge
transfer during one period to be nonzero?
Let the pumping potential $\phi({\bf r},t)$ be characterized by 
a finite set of functions ${\bf g}(t)=\{g_{\alpha}(t)\}$,
$\alpha=1,..m$, which are periodic with the same period $t_0$:
\begin{equation}
\phi({\bf r},t)=\phi({\bf r}, {\bf g}(t))=
 \sum_{\alpha}\phi_\alpha({\bf r})g_\alpha(t). 
\end{equation}
The time evolution of the set of functions ${\bf g}(t)$ represents a 
motion of a point in 
$m$ dimensional space ${\cal M}$. Due to periodicity of 
$\phi({\bf r}, t)$, 
the trajectory ${\cal C}$ of this point is closed. 
The above mentioned
currents lead to $Q\neq 0$, 
provided ${\cal C}$ encloses a finite area in ${\cal M}$. 
This requires that $m\geq 2$.

To calculate $Q$ we will use the Keldysh technique for the Green function
matrix equation[9]

\begin{eqnarray}
(i\partial_t - H_0 - \phi({\bf r}, t))\hat{G}({\bf r}, {\bf r}'; t, t')-
I_{e-ph}(\{\hat{G}\})
\nonumber \\
=\delta({\bf r}-{\bf r}')\delta(t-t'), 
\label{Keldysh}
\end{eqnarray}
where $\hat{G}$ is a $2\times 2$ matrix, 
$G_{11}=0$, $G_{12}=G^A$, $G_{21}=G^R$ and
$G_{22}=G^K$, with $G^{R, A, K}$ being the retarded, advanced and 
Keldysh Green functions respectively.
$H_0$ is the Hamiltonian for electrons which 
includes impurity scattering potentials, and $I_{e-ph}$ denotes the
electron-phonon collision integral.  
The solution of Eq. (2) can be expanded in term of
the changing rate of the external field $\dot{\phi}({\bf r}, t)$,
provided the time of the electron diffuses across the sample
is shorter than the period of the external potential $t_0$.
In general, $\hat{G}({\bf r}, {\bf r'}; t, t)$ 
depends on the value of the potential
$\phi({\bf r}, t')$ at all the previous time $t' \leq t$. 
However, in the first order adiabatic approximation  
$G({\bf r},{\bf r}';t,t)$ is determined only by 
the external potential and its first time derivative at time $t$, 
i.e.  $\hat{G}({\bf r},{\bf r}';t,t)=\hat{G}(\{ \phi({\bf r}, t) \},
\{\dot{\phi}({\bf r}, t)\})$.
The local time dependence in this approximation  
allows us to express
the charge transfer $Q_i$ per period $t_{0}$ along $i$th direction as 

\begin{equation}
Q_i=\frac{e}{L_i m}\int^{t_{0}}_0 dt
Tr\{ P_i G^K(t, t) \}=e\int_{{\cal C}}\omega^i_\alpha({\bf g}) dg_\alpha 
\end{equation}
where ${P_i}$ is $i$th component of the momentum operator,
$L_i$ is the
dimension of the sample along $ith$ direction, $i=x,y,z$,
$Tr$ means integration over the space coordinates, and

\begin{equation}
\omega^i_\alpha({\bf g})=\frac{\partial}{\partial\dot{g}_\alpha}
Tr\{\frac{P_i}{mL_i}G^K(t,t)\}.
\end{equation}
Eq. (4) valids only in the leading order in $\Omega=2\pi/t_0$.
We have introduced differential $1-form$ 
$\omega^{i}_\alpha({\bf g})$ on the m dimensional space ${\cal M}$. 
Using Stoke's theorem one can 
convert the $1-form$ integral along the trajectory ${\cal C}$ 
into the $2-form$ integral over any surface ${\cal S}$ spanning 
${\cal C}$,
 
\begin{eqnarray}
&&Q_i ={e}\int_{\cal S} dg_\alpha\wedge dg_\beta
\pi^i_{{\alpha} {\beta}}({\bf g}), 
\nonumber \\
&&\pi^i_{\alpha \beta}({\bf g})= 
\{\frac{\partial}{\partial g_\alpha}
\frac{\partial}{\partial\dot{g}_\beta}
-\frac{\partial}{\partial g_\beta}
\frac{\partial}{\partial\dot{g}_{\alpha}}\}
Tr\{\frac{P_i}{2mL_i}G^K(t, t)\}.
\end{eqnarray}
In an isolated quantum mechanical system, the 2-form
$\pi^i_{\alpha\beta}$   
corresponds to the generalized adiabatic curvature
of discrete eigenstates discussed in [3].
The wedge product is skew symmetric, i.e.
$dg_\alpha \wedge dg_\beta=
- dg_\beta \wedge dg_\alpha$. 
Note that $Q$ in Eq.(5)  
doesn't depend on the spanning surface.

In the zero order adiabatic approximation
Keldysh Geen function $G_\epsilon^K({\bf r}, {\bf r'}, t)$
can still be expressed through retarded and advanced ones 
$G^{R, A}_{\epsilon}({\bf r}, {\bf r}',t)$
and Fermi distribution function $n_F(\epsilon)$  
  
\begin{equation}
G_{\epsilon}^K=
(G^R_\epsilon -G^A_\epsilon)
[1- 2n_{F}({\epsilon})]
\end{equation}
where matrix elements of $\hat{G}_\epsilon$  
correspond to the {\em instantaneous}
hamiltonian $H_0+\phi({\bf r}, {\bf g})$,

\begin{equation}
\hat{G}_{\epsilon}({\bf r}, {\bf r}',t)=
\int dt' \hat{G}({\bf r}, {\bf r}'; t-\frac{t'}{2}, t+\frac{t'}{2})
e^{i\epsilon t'}
\end{equation}
and depend on time only through ${\bf g(t)}$.

According to Eqs.(2), (7), the first order correction
to the adiabatic approximation for $G^K$ can be written as
\begin{eqnarray}
\delta G_{\epsilon}^K({\bf r}, {\bf r}',t)=
i\int d{\bf r}''
\{ [\frac{1}{2}-n_F(\epsilon)](\Gamma_{RR}-\Gamma_{AA})\nonumber \\
-\frac{\partial n_F(\epsilon)}{\partial\epsilon}\Gamma_{RA}\}
\end{eqnarray}
where for any $p=(R, A)$, $q=(R, A)$,

\begin{eqnarray}
\Gamma_{pq}({\bf r}, {\bf r'}, {\bf r"}, t)=
2 G^p_\epsilon({\bf r}, {\bf r}'',t)\partial_t G^p_\epsilon({\bf r}'',
{\bf r}',t)\delta_{pq} \nonumber \\
+ G^p_\epsilon({\bf r}, {\bf r}'',t)\partial_t \phi({\bf r}'', t)
G^q_\epsilon({\bf r}'', {\bf r}',t)
\label{solution}
\end{eqnarray}
The contribution of the first two terms in Eq.(8) to $Q_i$ can be
neglected provided 
${L_i}$ is much bigger than the elastic mean free path $l$.
Substituting Eq.(8) into Eq.(5),
we obtain $\pi^i_{\alpha \beta}({\bf g})$, 
\begin{eqnarray}
\pi^i_{\alpha \beta}({\bf g})=
\frac{\partial \Xi_i(\beta, {\bf g})}{\partial g_\alpha(t)} 
-\frac{\partial \Xi_i(\alpha, {\bf g})}{\partial g_\beta(t)}, 
\nonumber \\
\Xi_i(\alpha, {\bf g})=
\int d{\bf r}d{\bf r_1} \phi_\alpha({\bf r_1})
\frac{1}{2mL_i}(\frac{\partial}{\partial r_i} - \frac{\partial}{\partial r'_i})
\nonumber \\
\int d\epsilon \frac{\partial n_F(\epsilon)}{\partial \epsilon} 
 G^R_\epsilon({\bf r}, {\bf r}_{1},t) 
G^A_\epsilon({\bf r}_1, {\bf r}',t)|_{{\bf r} \rightarrow {\bf r}'}. 
\end{eqnarray}

Due to the disorder, 
the charge $Q_i$ is a random sample-specific quantity. To characterize 
$Q_i$ we calculate its average $<Q_i>$ and variance
$<(\delta Q_i)^2>$($< >$ stands for the averaging over
realizations of the random potential).
In the following we assume that
$L_{\phi} \gg L_{z}, L_x, L_y \gg l$, where  
$L_{\phi}$ is the electron dephasing length. 
In this case one can express $<Q>$  and $<(\delta Q_i)^2>$ 
through $<\pi^i_{\alpha\beta}({\bf g})>$ and 
variance $\Pi^i_{{\alpha}{\beta}{\alpha'}\beta'}({\bf g}, {\bf g}')
= <\delta\pi^i_{\alpha \beta}({\bf g}) \delta \pi^i_{\alpha'\beta'}
({\bf g}')>$.
These quantities can be evaluated in a standard way\cite{abr}. 
Following Eq. (10), $<\pi^i_{\alpha\beta}>$ can be calculated in
terms of diagrams in Fig.1a,
\begin{figure} 
\begin{center}
\leavevmode
\epsfbox{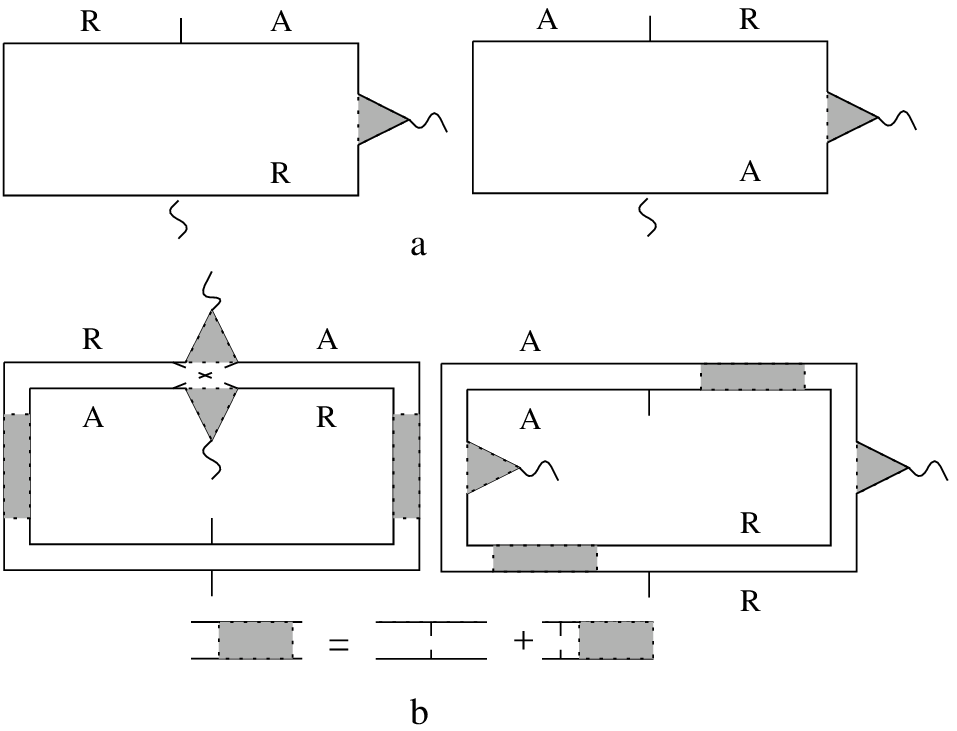}
\end{center} 

Fig.1.a).Diagrams for $<\pi^i_{\alpha\beta}({\bf g})>$; 
b).Diagrams for $\Pi^i_{\alpha\beta\alpha'\beta'}({\bf g}, {\bf g'})$. 
Solid lines represent retarded(R) and advanced(A) 
electron Green functions, 
dashed lines denote the impurity averaging, 
wavy lines represent the
external potential.
Shaded triangles stand for ${\cal D}({\bf r}, {\bf r}')$,
shaded boxes for diffusons ${\cal D}(\omega,
{\bf r}, {\bf r}';{\bf g}, {\bf g'})$.
In b)., the shaded triangles with two dashed lines
represent Hikami boxes.
\end{figure}

\begin{eqnarray}
<\pi^i_{\alpha\beta}({\bf g})>=\frac{\nu D_0}{2\mu_{F}L_i}
\int d{\bf r} d{\bf r'} \phi_\alpha({\bf r})\phi_\beta({\bf r'})
\frac{\partial{\cal D}({\bf r}, {\bf r}')}{\partial r_{i}}.
\end{eqnarray}
where $\nu$ is the mean density of states in the metal,
$\mu_F$ is the Fermi energy, $D_0$ is
the electron diffusion coefficient.
Note that $<\pi^i_{\alpha\beta}>$
doesn't depend on the external field, i.e. on ${\bf g}$ 
in the leading order of $\phi/\mu_F$.

$\Pi^i_{\alpha\beta\alpha'\beta'}$ is determined by 
diagrams in Fig.1b and can be presented in the form, 

\begin{eqnarray}
&&\Pi^i_{{\alpha}{\beta}{\alpha'}{\beta'}}
=\tilde{\Pi}^i_{{\alpha}{\beta}{\alpha'}{\beta'}}
+\tilde{\Pi}^i_{{\beta}{\alpha}{\beta'}\alpha'}
-\tilde{\Pi}^i_{{\alpha}{\beta}\beta'{\alpha'}}
-\tilde{\Pi}^i_{{\beta}{\alpha}{\alpha'}\beta'}
\nonumber\\
&&\tilde{\Pi}^i_{{\alpha}{\beta}{\alpha'}{\beta'}}({\bf g}, {\bf g}')
=\frac{D^2_0}{4L_i^2}\int d\epsilon d\omega 
\frac{\partial n_F(\epsilon+\omega/2)}{\partial \epsilon} 
\frac{\partial n_F(\epsilon-\omega/2)}{\partial \epsilon'} 
\nonumber \\
&&\int d{\bf r}_1 d{\bf r}'_1 
d{\bf r} d{\bf r'} 
\phi_{\alpha}({\bf r_1})\phi_{\alpha'}({\bf r'_1})
\nabla_{{\bf r}_1} 
{\cal D}({\bf r}_1, {\bf r})\cdot  
\frac{\partial^2\Lambda}{\partial g_\beta\partial g_{\beta'}}
\end{eqnarray}
which explicitly demonstrates antisymmetry
of the pumping perturbations. 
$\Lambda$ is given as
\begin{eqnarray}
\Lambda= Re [{\cal D}(\omega, {\bf r}, {\bf r'}; {\bf g}, {\bf g}')
{\cal D}(\omega, {\bf r'}, {\bf r}; {\bf g}, {\bf g}')]
\nabla_{{\bf r}'_1}
{\cal D}({\bf r}'_1, {\bf r'})\nonumber \\
+ {\cal D}(\omega, {\bf r}, {\bf r'}; {\bf g}, {\bf g}')
{\cal D}^*(\omega, {\bf r'}, {\bf r}; {\bf g}, {\bf g}')
\nabla_{{\bf r}'_1} {\cal D}({\bf r}'_1, {\bf r}). 
\end{eqnarray}
${\cal D}({\bf r},{\bf r}')$ and 
${\cal D}(\omega, {\bf r}, {\bf r}'; {\bf g}, {\bf g}')$ 
satisfy the following equations
\begin{equation}
D_0 \nabla^2 {\cal D}({\bf r}, {\bf r}')=\delta({\bf r}-{\bf r}')
\end{equation}
\begin{eqnarray}
\{i [\omega+ \delta \phi({\bf r})]
+ D_0\nabla^2 \} 
{\cal D}(\omega, {\bf r}, {\bf r}'; {\bf g}, {\bf g'})
=\delta({\bf r} -{\bf r}'),
\end{eqnarray}
where $\delta\phi({\bf r}, {\bf g}, {\bf g'})  
=\phi({\bf r}, {\bf g'}) - \phi({\bf r}, {\bf g'})$. 
For Eqs.(14), (15),
we use usual boundary conditions: 
${\cal D}({\bf r},{\bf r}')={\cal D}(\epsilon-\epsilon',{\bf r},{\bf r}')=0$ 
when ${\bf r}$ or ${\bf r}'$ is at open boundaries[11,12].

Let us now consider the sample sketched in Fig.2 
with two gates(labled by $\alpha=1,2$),
biased with a.c. voltages of the same frequency and with a phase shift
$\delta=\delta_{1}-\delta_{2}$,
\begin{equation}
V_{\alpha}(t)=V_0\times\sin(\Omega t+\delta_{\alpha}).
\end{equation}
In this case, $m=2$.  
Let us assume that the potential induced in the metal  
by the voltages $V_{\alpha}$ is
screened with a screening length $r_0$ much less than $L_x$ and
\begin{eqnarray}
&&g_\alpha(t)=\sin(\Omega t +\delta_\alpha), 
\nonumber \\
&&\phi_\alpha({\bf r})= 
\frac{CV_0 r_0}{L_yW}\theta({r_{0}}-x) 
\theta(\frac{W^2}{4}-(Z_\alpha -z)^2),
\end{eqnarray}
where $C$ is the capacitance of the gate, 
$W \gg r_0$ is the width of the gate along
$z$ direction, $Z_{1,2}$ are the $z$ coordinates of the center of the
gate 1,2; and $\theta(z)$ is the step function: $\theta(z \geq 0)=1$ 
while $\theta(z<0)=0$.
\begin{figure}
\begin{center}
\leavevmode
\epsfbox{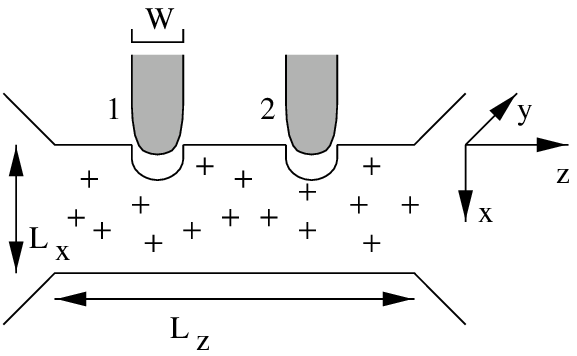}
\end{center} 
Fig.2. Geometry of the sample. Shaded bars represent gates 1 and 2, 
crosses represent random scatters. 
\end{figure}

To evaluate $<Q_z>$ in the leading order in $V_0/\mu_F \ll 1$, one has to
substitute Eqs.(11), (17) into Eqs.(5)
and take into account that
\begin{equation}
\int_{\cal S} dg_\alpha \wedge dg_\beta=
\int_{0}^{t_{0}} dt \{\partial_t g_\alpha(t) g_\beta(t)
-\partial_t g_\beta(t) g_\alpha(t) \}.
\end{equation}
Given the volume of the sample $v=L_xL_yL_z$ and the total
number of electrons inside the sample $N$
we present $<Q_z>$ as
\begin{equation}
<Q_z>=f_{0} e\sin(\delta) 
(\frac{C V_0 r^2_{0} }{v\mu_F })^{2}N.
\end{equation}
$f_0 \sim 1$ is a geometry dependent factor.  

To determine $\Pi^i_{\alpha\beta\alpha'\beta'}$ one has to 
solve Eq. (15). When 

\begin{equation}
\frac{C V_0 r^2_0}{v} \ll E_T=\frac{D_0}{L_z^2},
\end{equation}
this can be done by 
using  perturbation theory with respect to $g_\alpha(t)$.
Keeping only the bilinear
in ${\bf g}$ and ${\bf g'}$ contributions to
${\cal D}(\omega,{\bf r}, {\bf r}'; {\bf g}, {\bf g'})$,
we express the solution of Eq.(15) in terms of 
${\cal D}^0(\omega, z, z')$,
the ${\bf g}$ independent solution of Eq. (15) with $\phi({\bf r}, {\bf g})=0$.
As usual,
in the quasi-one dimensional case, we neglected $x, y$ 
dependences of ${\cal D}^0$.
Using Eq.(12), we 
then express $\Pi^z_{\alpha\beta\alpha'\beta'}({\bf g}, {\bf g'})$
in terms of ${\cal D}^0$.
It is important to notice that 
$\Pi^z_{\alpha\beta\alpha'\beta'}$ is
independent of ${\bf g}, {\bf g'}$
in the leading order of $CV_0 r_0^2/vE_T \ll 1$. 
Thus, according to Eq.(5), $\sqrt{<(\delta Q_z)^2>}$ is proportional to
the area $S$ enclosed by the trajectory ${\cal C}$. 
As a result for $CV_0 r_0^2/v \ll E_T$, 
\begin{equation}
<(\delta Q_z)^2>=(e\sin\delta)^{2}(\frac{CV_0 r^2_0}{v
E_{T}})^{4} f_1(\frac{T}{E_T}).
\end{equation}
here the functions $f_1(\eta)$ has the following asymptotics,
\begin{equation}
f_{1}(\eta) \propto \left \{ \begin{array}{cc}
1, &\mbox{$ \eta \ll 1$} \\
{\eta^{-1}}, &\mbox{$\eta \gg 1$}
\end{array}
\right.
\end{equation}
In the limit $CV_0 r_0^2/v \gg E_T$, 
$\pi^z_{{\alpha} {\beta}}
({\bf g})$ is a random quantity
in 2-D space ${\cal M}$ of $\{{\bf g}\}$ with 
the "correlation length" $|\delta {\bf g}_c| \sim E_T v/CV_0 r_0^2$
(which is much less than unity).
$\sqrt{<(\delta Q_z)^2>}$ is determined by the amount 
of "flux" of random $\pi^z_{\alpha\beta}({\bf g})$
field that threads the loop ${\cal C}$ and
should increase slower than the enclosed area $S$ itself. 
This limit will be considered elsewhere. 

It follows from Eqs.(19), (21)  
that the standard deviation of $Q_z$ is much larger than  
its average, provided $N$ is much bigger than ${G_z}^2$, 
($G_z$ is the dimensionless conductance in the $z$ direction)and
the amplitude of the external perturbation is not too large.
In this case the value of $Q_z$ is entirely determined by quantum interference 
effects.
In the opposite limit, $Q_{z}$ can be characterized by its average given
by Eq.(19).

According to Eq.(19), $<Q_z>$ is proportional to
$\sin\delta$, i.e. this quantity
changes sign together with $\delta$ and vanishes
at $\delta=0$. 
In fact this is as well correct for the charge transfer $Q_z$
in a specific sample of given realization of disorders,
in the weak perturbation limit. 
One can see this from Eq.(5), taking into account that
$\pi^i_{\alpha\beta}({\bf g})$ is a random
quantity independent of ${\bf g}$ when ${\bf g}$ is small. 
More generally, 
the charge transfer $Q$ changes sign when $\delta
\rightarrow -\delta$ for arbitrary amplitude of
the external perturbation although  
the simple $\delta$ dependence
in Eq. (19), (21) is valid only in the weak external field.
Indeed, T-invariance requires that changing the direction
of the trajectory ${\cal C}$ in the space ${\cal M}$ (which
corresponds to changing sign of $\delta$ 
for the case of two gates) 
should result in the change of sign of the charge:
$Q_{\hookleftarrow}=-Q_{\hookrightarrow}$ where 
$\hookleftarrow$($\hookrightarrow$)
corresponds to clockwise(counterclockwise) motion
along the same closed trajectory ${\cal C}$.  
In the presence of magnetic field, this identity acquires a form
\begin{equation}
Q_{\hookleftarrow}({\bf H})=-Q_{\hookrightarrow}(-{\bf H}).
\end{equation}
Here $Q_{\hookleftarrow}({\bf H})$
and $Q_{\hookrightarrow}({\bf H})$ are charges which correspond to
clockwise and counterclockwise motion
along the same closed trajectory ${\cal C}$ in the space 
${\cal M}$ and ${\bf H}$ is the external magnetic field.
When the external potential is weak, one can neglect
the ${\bf g}$-dependence of $\pi^i_{\alpha\beta}$ field.
In this limit, $Q$ is proportional to the area $S$  
enclosed by trajectory ${\cal C}$
and $\pi^{i}_{\alpha\beta}({\bf g}\approx 0)$.
Thus, Eq.(23) indicates that in the vicinity of ${\bf g}=0$ 
$\pi^i_{\alpha\beta}({\bf g})$ is an even function of ${\bf H}$.
At small amplitudes of oscillations of the external
potential, this leads to
$Q_{\hookleftarrow (\hookrightarrow)}({\bf H})=Q_{\hookleftarrow
(\hookrightarrow)}(-{\bf H})$,
provided without magnetic fields the system occupies a 
T-invariant state.

As it is usual in mesoscopic physics, the magnetic field dependence of
$Q_{z}({H})$
exhibits random sample specific fluctuations with a characteristic
period $\Delta H\sim {\Phi_{0}}/{L_zL_x}$. Here $\Phi_{0}$ is the flux
quanta (the magnetic field is applied along $y$ direction).
At high temperatures, when the dephasing length is shorter than 
the sample size, mesoscopic effects
become exponentially small and $Q_{z}$ is determined by
Eq.(19).

Different mesoscopic mechanism of the adiabatic charge transport has been
discussed in[13].
Inelastic electron-phonon processes cause shifts of 
the center of mass of the electron wave functions and thus  
contribute to $Q_i$. However, in the case of open geometry
samples, this contribution is small compared with Eq.(20) as 
$(\tau_{e-ph} E_T)^{-1}\ll 1$, where $\tau_{e-ph}$ is the 
electron-phonon inelastic scattering time.

In conclusion, we would like to stress that the d.c. current 
discussed above is proportional to the frequency of the oscillations
of the external potential $\Omega$. This distinguishes the considered
above effect from the usual photovoltaic effect.
The latter effect in the low frequency limit is dominated 
by the relaxations of nonequilibrium distribution of electrons
in the presence of external field via electron-phonon inelastic processes
[13].
Such a mechanism leads to a randomly directed d.c. current 
proportional to the absorption rate of the external field
even in the presence of a {\em single} pumping gate voltage. 
It means that the photovoltaic d.c. current is proportional to $\Omega^2$,
as the frequency dependence of the absorption rate
of the external field is.
In an open sample in the low frequency limit,
the photovoltaic current is smaller than the adiabatic
current by a factor $\Omega/E_T \ll 1$.

It should be emphasized that the value of
$Q$ in a finite open mesoscopic system is 
{\em not quantized}. An approximate
quantization of $Q$ can be achieved in
a Coulomb blockade regime, for a pumping of the charge through
a quantum dot, which is weakly connected with the source 
and drain[14,15,16]. $Q$ turns out to be
rather well quantized provided the dimensionless conductance of the 
device is small. However, under these conditions,
pumping is an entirely classical effect. It is not related to the
quantum interference mechanism discussed in this paper.

While preparing the manuscript we have learned about the
paper of P. Brower where similar results were also obtained.
We acknowledge useful discussions with I. Aleiner,
C. Marcus. 
Work of B. Spivak is supported by Division of material science, U.S. National
Science Foundation under Contract No DMR-9205114.  
Work of F. Zhou is supported by ARO under contract
DAAG 55-98-1-0270.

\end{multicols}

\end{document}